\documentclass[fleqn]{revtex4-1}

\usepackage{graphicx}
\usepackage{amsmath, amsfonts, amssymb, amsbsy}
\usepackage{MnSymbol}
\usepackage[usenames,dvipsnames]{xcolor}
\usepackage{marvosym}
\usepackage{sidecap}
\usepackage{ifsym}
\usepackage{color}
\usepackage{ulem}

\newcommand{\beq} {\begin{equation}}
\newcommand{\eeq} {\end{equation}}

\newcommand{\tc} {{T_{\rm c}}}
\newcommand{\lsco} {{La$_{2-x}$Sr$_x$CuO$_4$}} 
\newcommand{\ybco} {{YBa$_{2}$Cu$_3$O$_{6+y}$ }}

\begin{document}
\begin{center}

\textbf{\large New Perspective on the Phase Diagram of Cuprate High-Temperature Superconductors}\\
\vspace{3cm}

{\bf \flushleft Damian Rybicki$^*$}\\
{University of Leipzig, Faculty of Physics and Earth Sciences,  Linn\'estr. 5, 04103 Leipzig, Germany}, and\\
{AGH University of Science and Technology, Faculty of Physics and Applied Computer Science,
Department of Solid State Physics, al. A. Mickiewicza 30, 30-059 Krakow, Poland}\\
Email: ryba@agh.edu.pl\\
Tel: +48.12.6172946

{\bf \flushleft Michael Jurkutat}\\
{University of Leipzig, Faculty of Physics and Earth Sciences,  Linn\'estr. 5, 04103 Leipzig, Germany}\\
Email: jurkutat@physik.uni-leipzig.de\\
Tel: +49.341.9732605\\

{\bf \flushleft Steven Reichardt}\\
{University of Leipzig, Faculty of Physics and Earth Sciences,  Linn\'estr. 5, 04103 Leipzig, Germany}\\
Email: steven.reichardt@uni-leipzig.de\\
Tel: +49.341.9732605

{\bf \flushleft Czes\l aw Kapusta}\\ 
AGH University of Science and Technology, Faculty of Physics and Applied Computer Science,
Department of Solid State Physics, al. A. Mickiewicza 30, 30-059 Krakow, Poland\\
Email: kapusta@agh.edu.pl\\
Tel: +48.12.6172554\\

{\bf \flushleft J\"urgen Haase}\\
{University of Leipzig, Faculty of Physics and Earth Sciences,  Linn\'estr. 5, 04103 Leipzig, Germany}\\
Email: j.haase@physik.uni-leipzig.de\\
Tel: +49.341.9732601\\
\end{center}
\newpage

\textbf{Universal scaling laws can guide the understanding of new phenomena, and for cuprate high-temperature superconductivity such an early influential relation showed that the critical temperature of superconductivity ($\tc$) correlates with the density of the superfluid measured at low temperatures. This famous Uemura relation has been inspiring the community ever since. Here we show that the charge content of the bonding orbitals of copper and oxygen in the ubiquitous CuO$_2$ plane, accessible with nuclear magnetic resonance (NMR), is tied to the Uemura scaling. 
This charge distribution between copper and oxygen varies between cuprate families and with doping, and it allows us to draw a new phase diagram that has different families sorted with respect to their maximum $\tc$. Moreover, it also shows that $\tc$ could be raised substantially if we were able to synthesize materials in which more oxygen charge is transferred to the approximately half filled copper orbital.}\\

\begin{figure}
\includegraphics[width=0.99\textwidth]{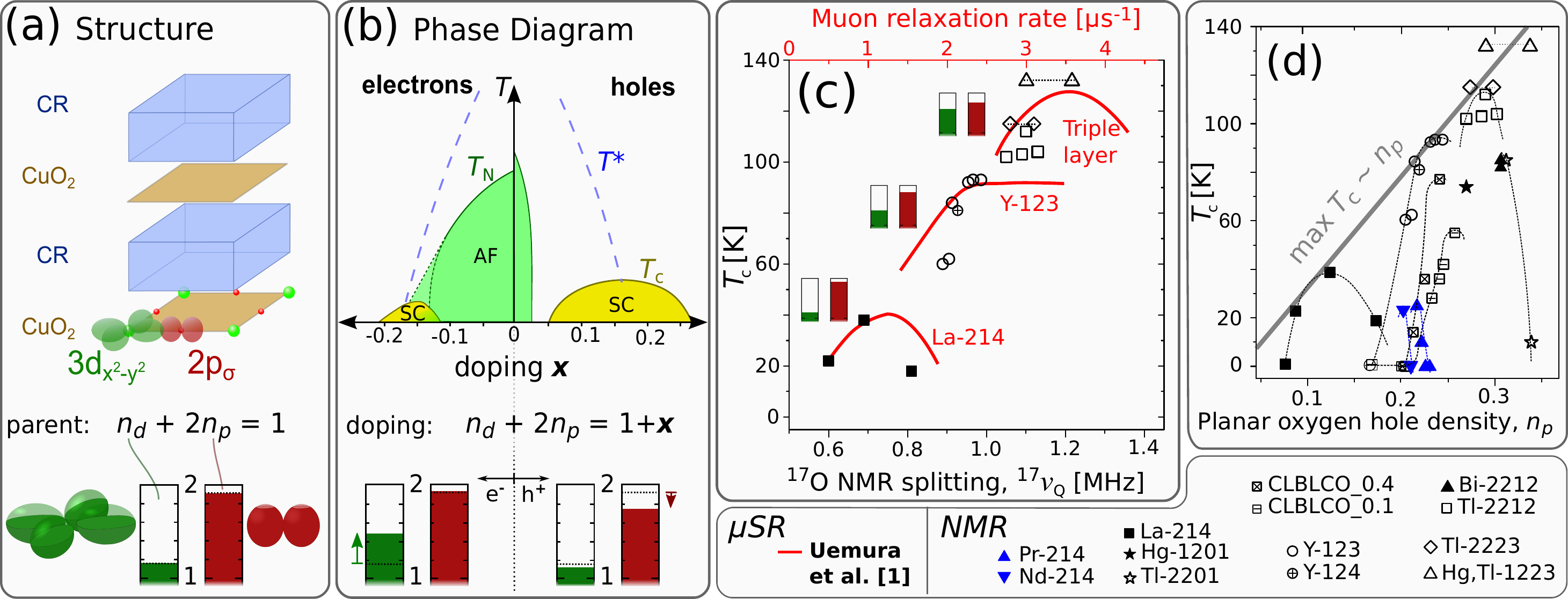}
\caption{\label{Uemura}(a) \textit{top:} the cuprates  layered structure has CuO$_2$ planes and charge reservoir (CR) layers; (a) \textit{bottom:} the bonding orbitals in the CuO$_2$ plane, i.e., Cu 3d$_{x^2-y^2}$ and O 2p$_\sigma$, share the nominal Cu $3d$ hole of the Cu$^{2+}$ ion (indicated filling measured with NMR);
(b) \textit{top:} schematic electronic phase diagram of the cuprates for electron (\textit{left}) and hole (\textit{right}) doping $x$, with antiferromagnetic (AF) and superconducting (SC) phases; at low doping the pseudogap reigns below $T^*$; (b) \textit{bottom:} doped electrons go to the 3d$_{x^2-y^2}$ orbital almost exclusively, while doped holes predominantly go to the 2p$_\sigma$ orbital;
(c) \textit{solid red:} Uemura plot \cite{Uemura1989}, i.e., $\tc$ vs. muon spin relaxation rate (upper abscissa); \textit{black symbols:} $\tc$ vs. planar oxygen quadrupole splitting $^{17}\nu_\mathrm{Q}$ (lower abscissa). For triple layer Tl-2223 and Hg,Tl-1223 the pairs connected with a dotted line belong to the same sample and correspond to planar O sites of inner and outer layer (smaller splitting corresponds to underdoped inner CuO$_2$ layer). 
(d) $T_\mathrm{c}$ vs. planar O hole density $n_p$ calculated from $^{17}\nu_\mathrm{Q}$ for all available data (see text). }
\end{figure}

\section{Introduction}
The cuprates' essential building blocks are the CuO$_2$ plane and charge reservoir  layers that separate the planes, cf. Fig.~1(a). While the square-planar CuO$_2$ plane with a Cu $3d_{x^2-y^2}$ orbital bonding to four O $2p_\sigma$ orbitals is quite universal, the charge reservoir chemistry can vary widely. The antiferromagnetic parent compound can be doped with holes or electrons by alteration of the charge reservoir layers so that the static magnetism vanishes and new electronic phases emerge, cf. Fig.~1(b).

The original plot by Uemura et al. \cite{Uemura1989}~depicted in Fig.~1(c) (in red) shows $\tc$ correlated with the muon spin relaxation rate $\sigma_0$ (extrapolated to $T$ = 0 K), which is proportional to the superfluid density divided by the effective mass ($\sigma_0 \propto n_s/m^*$). 
This relation holds for the underdoped materials and nicely orders different cuprate families. This and subsequent scaling laws have remained stimulating up to now, and some were shown to be valid for other superconductors as well \cite{Dordevic2002, Savici2002, Homes2004, Tallon2006, Homes2009, Dordevic2013, Wu2010, Homes2012, Shengelaya2015}. 
Also shown in Fig.~1(c) (in black) is the planar $^{17}$O NMR quadrupole splitting ($^{17}\nu_\mathrm{Q}$) that measures the O 2$p_\sigma$ hole content. The resemblance of this temperature independent charge density at the planar oxygen, set by material chemistry, and $\sigma_0$,  the density of the superfluid at very low temperatures, is striking.

\section{Results and Discussion}
Nuclear magnetic resonance (NMR) as a versatile local bulk probe revealed various trends among certain parameters for the cuprates. \cite{Slichter2007} However, NMR spin shifts, nuclear relaxation, or local electric field gradients  do not lend themselves easily  to simple physical pictures. 
For example, early work related $\tc$ to the planar O and Cu splittings \cite{Zheng1995a}, but it did not attract as much interest as the Uemura plot. Subsequent work \cite{Haase2004} showed that the splittings are due to the hole content of the Cu 3$d_{x^2-y^2}$  ($n_d$) and O 2$p_\sigma$  ($n_p$) orbitals, and that they measure the chemical hole-doping $x$ (as in La$_{2-x}$Sr$_x$CuO$_4$), i.e., $x=\Delta n_d +2 \Delta n_p$. Here, $n_d$ and $n_p$ are measured with NMR based on calibrating the quadrupole splittings with atomic spectroscopy data, in contrast to the earlier model \cite{Zheng1995a}.  
Very recently \cite{Jurkutat2014}, it was found that even electron-doping is quantitatively accounted for with $n_d$ and $n_p$ from NMR. In addition, it was found that the parent materials differ substantially in $n_d$ and $n_p$, however, the relation $ n_d+2 n_p =1$ is obeyed, i.e., the Cu and O bonding orbitals in the CuO$_2$ plane share the nominal d-shell Cu$^{2+}$ hole differently. This results in the sorting of the families as shown in Fig.~1(d), and one recognizes that a large $n_p$ is a prerequisite for a high maximum $\tc$ (i.e. for optimal doping). 

\begin{figure}
\includegraphics[width=0.45\textwidth]{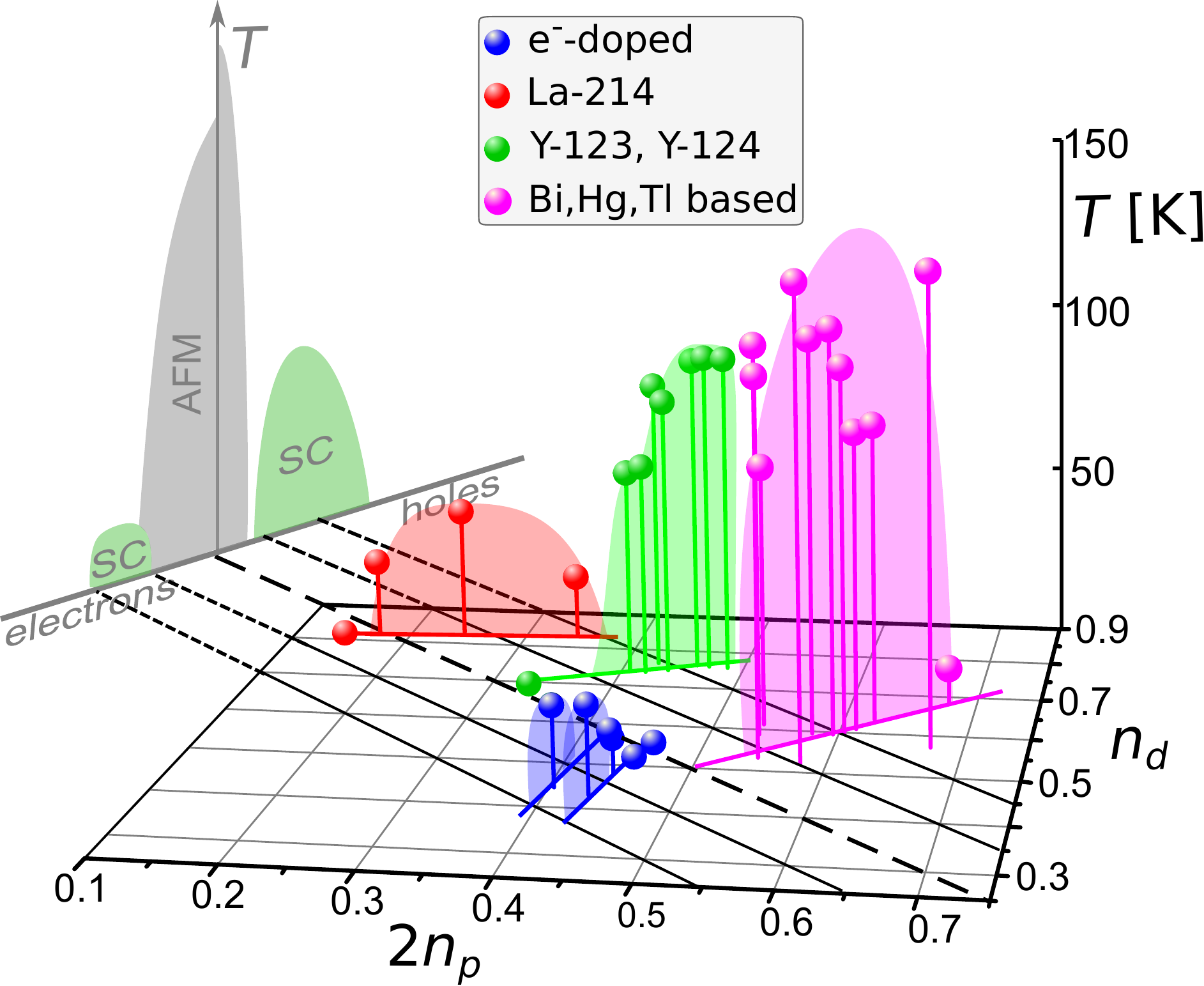}
\caption{\label{3d}(Color online) $\tc$ as a function of oxygen (2$n_p$) and copper ($n_d$) hole content for electron-doped Pr-214 and Nd-214, hole-doped La-214, Y-123 and Bi-, Hg-, Tl-based compounds. The commonly used phase diagram ($T$ vs $x$) appears as a projection (upper left). Black dashed bold line is for the undoped case ($x$=0), thin black lines correspond to doping $x$ changing with a step of 0.1.}
\end{figure}

While the knowledge of $^{17}\nu_\mathrm{Q}$ is sufficient for calculating $n_p$,  determination of $n_d$ requires the splittings measured at both nuclei \cite{Haase2004, Jurkutat2014}. 
Since NMR can only measure $^{17}$O enriched samples, the number of reports on $^{17}\nu_\mathrm{Q}$ is much lower compared with $^{63}\nu_\mathrm{Q}$ (see Supplementary). 
Therefore, we could convert all the planar oxygen splittings from the literature, but only the corresponding subset of copper splittings. 
Although the plot of $\tc$ vs. $n_p$ in Fig.~1(d) is similar to Fig.~1(c), it additionally includes non-superconducting and overdoped compounds. 
We also recognize in Fig.~1(d) a parabolic-like dependence of $\tc$ on the oxygen charge $n_p$, which resembles the typical phase diagram that shows a dome-like dependence of $\tc$ on $x$. 
Since there is no superfluid in non-superconducting materials (parents, and for doping outside the $\tc$ dome), they cannot be shown in the Uemura plot. 
Furthermore, the correlation between $\sigma_0$ and $^{17} \nu_\mathrm{Q}$ is lost in the overdoped regime where $\sigma_0$ decreases with increasing doping \cite{Uemura1993, Niedermayer1993} (which was attributed to a decrease of $n_s$ \cite{Tallon1997}).

In Fig.~1(d) we also included results for the electron doped materials that we have obtained very recently.\cite{Jurkutat2014} 
For electron-doped Nd$_{1.85}$Ce$_{0.15}$CuO$_4$ the muon relaxation rate and the superfluid density were reported to be very similar to that of hole-doped YBa$_2$Cu$_3$O$_{6+y}$. \cite{Shengelaya2005, Homes1997, Homes2004}
We find that $\mu$SR data for electron doped compounds are also in agreement with $^{17}\nu_\mathrm{Q}$ splittings (see Supplementary) and corresponding hole contents for those families, cf. Fig.~1(d).
Electron-doping appears to be less efficient in providing a high $\tc$, but the rather high oxygen hole content of the parent materials Pr(Nd)$_2$CuO$_4$ suggests that hole-doping should result in much higher $\tc$. 
Clearly, a large $n_p$ is only a prerequisite for a high $\tc$, but is not sufficient, as expected for such a material chemistry parameter. 
We also do not know whether this empirical relation remains valid for higher oxygen hole content. 
If it does, the $\tc$ of the cuprates might be raised by the proper chemistry substantially (we estimate 300 K to 400 K per oxygen hole from the straight line in Fig.~1(d)).
Since the charge transfer between Cu and O is governed by $1=n_d+2 n_p$ for the parent materials, we also conclude that compounds with the highest $\tc$ favor a smaller Cu hole content. 

These findings suggest a different kind of cuprate phase diagram that we present in Fig.~2. 
It does not use the average doping ($x$) as abscissa, but distinguishes between the oxygen and copper charges. The ordinary phase diagram ($T$ vs. $x$), cf. Fig.~1(b), appears as a projection that has $x=n_d+2 n_p -1$ on its abscissa. 
It must be of great interest to discuss the various cuprate properties probed by other methods in terms of the new phase diagram, however, this task is beyond the scope of this manuscript, and we would like to raise only a few ideas here. 

As to the parent compounds, while it is of interest to understand how the exchange coupling $J$ varies as a function of $n_d$ and $n_p$, the exchange between the CuO$_2$ planes and with it the Ne\'el temperature will depend on the charge reservoir layers, and correspondingly, $T_{\rm N}$ shows no simple trend. The parent materials of the electron-doped cuprates promise a large $\tc$ upon \textit{hole} doping. 

Doping appears essential for unlocking the maximum $\tc$, and it changes $n_d$ and $n_p$ in a family-specific manner. 
While hole-doping changes mostly $n_p$, electron-doping almost exclusively affects $n_d$. 
According to our analysis, all families show optimal doping near $x= n_d +2 n_p-1\approx \pm 0.15 $. 
This suggests that optimal doping is related to the parent magnetism rather than the distribution of charges between Cu and O.
Electron-doping is less effective in unlocking the maximum $\tc$. 
The hole-doped compounds appear in three separate groups: (1) \lsco, (2) \ybco and other cuprates of that structure, as well as YBa$_2$Cu$_4$O$_8$, and (3) Bi, Tl and Hg based families, which have the highest $\tc$ values.  

Another important issue concerns the heterogeneity of the cuprates. 
We know from  NMR that the static charge and spin density can vary drastically within the CuO$_2$ plane, in particular between different cuprate families \cite{Haase2003}. 
For example, the charge density in terms of the total doping $x$ may easily vary by $\Delta x \approx 0.05$\cite{Rybicki2009, Jurkutat2013, Singer2002}. Since
$\tc$ is not in a simple relation to this static inhomogeneity, only the average $n_d$ and $n_p$ appear to matter. 
From this, one would conclude that inhomogeneity is either not important for the maximum $\tc{}$, or it is ubiquitous and dynamically averaged for NMR, depending on the chemical environment. \cite{Haase2003} 

Pressure has profound effects on $\tc$ and probably on $n_d$ and $n_p$. This would be very revealing, and some of us are engaged with new  high-pressure NMR experiments \cite{Meissner2011} and pursue this issue currently.

Concerning the electronic fluid: it is beyond doubt now that the susceptibility of a single electronic spin component cannot explain the cuprate NMR shifts \cite{Haase2009, Haase2012}. Instead, one needs at least a Fermi-liquid-like spin component that has a temperature-independent spin polarization above $\tc$, and a pseudogap-like spin component that is temperature-dependent far above $\tc$ for lower doping levels. A third, doping-dependent NMR shift term was recently identified \cite{Rybicki2015}, and it may represent the expected coupling between the two components. Therefore, it will be of great interest to see how the different spin components vary across the new phase diagram. 

To conclude, NMR measures the charge distribution in the bonding orbitals in the CuO$_2$ plane quantitatively, and since it reproduces the Uemura plot,  i.e., it finds the same ordering of families with respect to their maximum $\tc$, we now have  material chemistry parameters that are responsible for setting the highest $\tc$ and superfluid density. These findings inspired a new perspective on the cuprate phase diagram and it is very likely that the complex cuprate properties might be better understood when discussed in the context of the charge distribution in the CuO$_2$ plane.

\begin{acknowledgments}
 
\textbf{Acknowledgement} We are thankful to O.P. Sushkov, C. P. Slichter, G. V. M. Williams for helpful discussions, and acknowledge financial support by the University of Leipzig, the DFG within the Graduate School Build-MoNa, the European Social Fund (ESF) and the Free State of Saxony.
 
\end{acknowledgments}

\bibliography{Rybicki2013a}{}
\bibliographystyle{apsrev4-1}

\end{document}


\begin{center}

\textbf{\large Supplement to: New Perspective on the Phase Diagram of Cuprate High-Temperature Superconductors}\\
\vspace{3cm}

{\bf \flushleft Damian Rybicki$^*$}\\
{University of Leipzig, Faculty of Physics and Earth Sciences,  Linnestr. 5, 04103 Leipzig, Germany}, and\\
{AGH University of Science and Technology, Faculty of Physics and Applied Computer Science,
Department of Solid State Physics, al. A. Mickiewicza 30, 30-059 Krakow, Poland}\\
Email: ryba@agh.edu.pl\\
Tel: +48.12.6172946

{\bf \flushleft Michael Jurkutat}\\
{University of Leipzig, Faculty of Physics and Earth Sciences,  Linnestr. 5, 04103 Leipzig, Germany}\\
Email: jurkutat@physik.uni-leipzig.de\\
Tel: +49.341.9732605\\

{\bf \flushleft Steven Reichardt}\\
{University of Leipzig, Faculty of Physics and Earth Sciences,  Linnestr. 5, 04103 Leipzig, Germany}\\
Email: steven.reichardt@uni-leipzig.de\\
Tel: +49.341.9732605

{\bf \flushleft Czes\l aw Kapusta}\\ 
AGH University of Science and Technology, Faculty of Physics and Applied Computer Science,
Department of Solid State Physics, al. A. Mickiewicza 30, 30-059 Krakow, Poland\\
Email: kapusta@agh.edu.pl\\
Tel: +48.12.6172554\\

{\bf \flushleft J\"urgen Haase}\\
{University of Leipzig, Faculty of Physics and Earth Sciences,  Linnestr. 5, 04103 Leipzig, Germany}\\
Email: j.haase@physik.uni-leipzig.de\\
Tel: +49.341.9732601\\
\end{center} 
\newpage

\begin{figure}
\includegraphics[width=1\textwidth]{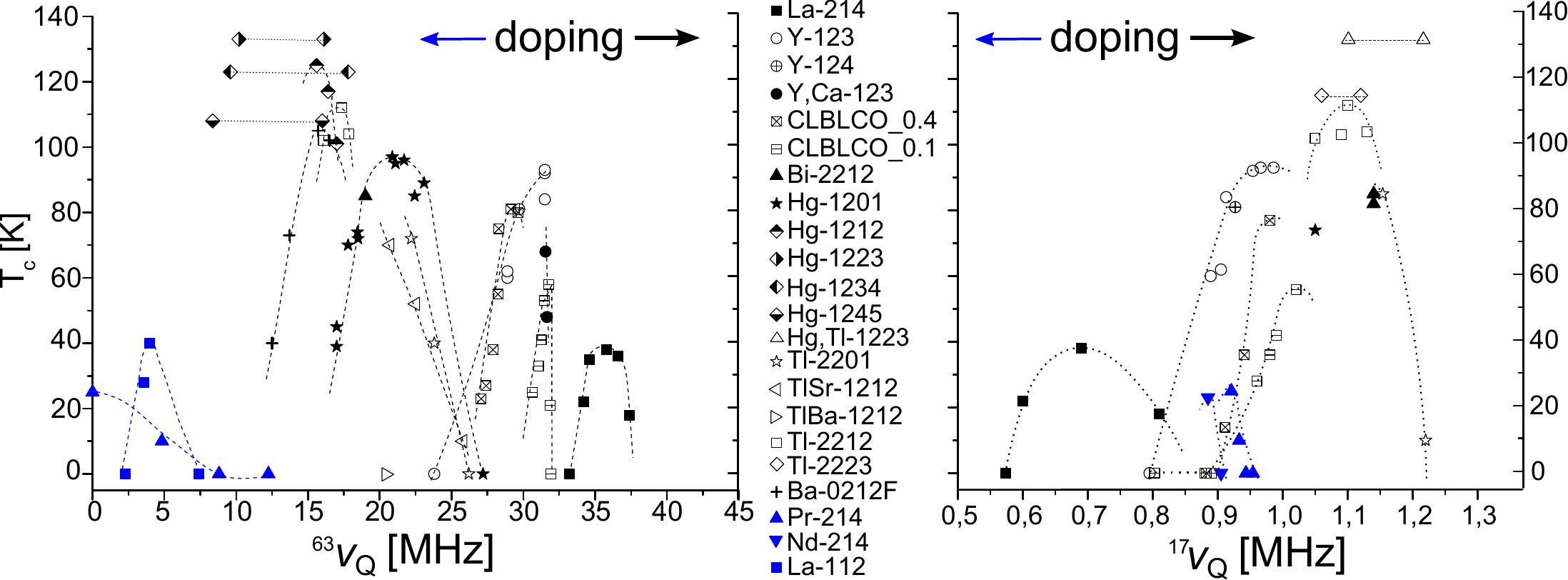}
\caption{(Color online) Phase diagrams of cuprates based on quadrupole splittings of the planar copper ({\it left panel}) and oxygen ({\it right panel}). Black and blue symbols and arrows (indicating increase of doping  $x$) are for hole and electron doped systems, respectively. Dotted lines are guides to the eye and connect different doping levels for one family. 
For compounds with three or more CuO$_2$ layers there are two different planar Cu and O sites, which have different quadrupole splittings, hence the horizontally connected data points belong to the same sample.}
\label{fig_O} 
\end{figure}

A detailed analysis of the planar oxygen and copper electric field gradient tensors in the cuprates is presented elsewhere \cite{Rybicki2014}. Here we only present collected literature values of the quadrupole splittings of planar copper $\left(^{63}\nu_\mathrm{Q}\right)$ and oxygen ($^{17}\nu_\mathrm{Q}$) (table 1 and 2, respectively).
All $^{17}$O and most of $^{63}$Cu splittings are also plotted in Fig.~\ref{fig_O}. In the case of the CLBLCO family there are $^{63}\nu_\mathrm{Q}$ data for more dopings.\cite{Keren2006} Since all show the same behaviour in the tables we show only two families, for which both $^{63}\nu_\mathrm{Q}$ and $^{17}\nu_\mathrm{Q}$ are available. 
There are also $^{63}\nu_\mathrm{Q}$ results for cuprates containing Bi, e.g. for single layer Bi-2201 family.\cite{Kato1997} 
However, the $^{63}$Cu NQR spectra are very broad with at least four peaks (ranging from 27 MHz to 35 MHz), despite the fact that all Cu sites are equivalent in the ideal Bi-2201 structure without the modulation in the Bi-O layer. 
Very broad $^{63}$Cu NQR spectra are a general feature of cuprates with this type of modulation \cite{Lipinski1990}, and we only quote one $^{63}\nu_\mathrm{Q}$ obtained from NMR on Bi-2212 single crystals,\cite{Ishida1994} for which also $^{17}\nu_\mathrm{Q}$ is available.

The hole contents of the oxygen $2p_\sigma$ ($n_p$) and copper $3d_{x^2-y^2}$ ($n_d$) orbitals, which are shown in the main paper, are calculated using the following formulas \cite{Jurkutat2014}:
\begin{align*}
n_p&=\frac{^{17}\nu_\mathrm{Q} - 0.39~\mathrm{MHz}}{2.45~\mathrm{MHz}} \\ 
n_d&=\frac{^{63}\nu_\mathrm{Q} +5.65(8-4n_p)~\mathrm{MHz}}{94.3~\mathrm{MHz}}.
\end{align*}

\newpage

 \begin{table}
 \caption{$^{63}$Cu quadrupole splitting $^{63}\nu_\mathrm{Q}$ for various families of cuprates in order of increasing doping ("und." - underdoped, "opt." - close to optimal doping, "ovd." - overdoped). For Tl- and Hg- families with three or more CuO$_2$ layers there are two Cu sites, the inner layer with Cu(1) and the outer layer Cu(2). 
 \label{EFG-data}}
\small
  \begin{tabular}{| c | c | c || c | c | c |  }
\hline
Family		&	$T_{\rm c}$ [K]		&	$^{63}\nu_\mathrm{Q} $  [MHz] 	&      Family		&	$T_{\rm c}$ [K] 		&	$^{63}\nu_\mathrm{Q}$ [MHz] \\   \hline \hline
La-214 ref.\cite{haase_planar_2004}		&	0	&	33.2		&  Bi-2212	&			86 (opt)  ref.\cite{Ishida1994}	&	19.0 \\ \cline{4-6}
  	&		22	&	34.2	& Tl-2201 ref.\cite{Fujiwara1991}	&			72 (ovd)	&	22.2\\
	&		35	&	34.6	&	&			40 (ovd)	&	23.8			\\
	&		38	&	35.8	&	&			0 (ovd	&	26.2	\\ \cline{4-6}
	&		36	&	36.6	& TlSr-1212 ref.\cite{Magishi1996b}	&			70 (ovd)	&	20.7\\ 
	&		18	&	37.4	& &			52 (ovd)	&	22.5	\\  \cline{1-3}
Y-123 ref.\cite{haase_planar_2004}&		0	&	23.8		&&			10 (ovd)	&	25.8	\\ \cline{4-6}
	&		60	&	28.9		& TlBa-1212 ref.\cite{Goto1996}	& 0 & 20.44 \\ \cline{4-6}
	&		62	&	28.9		&	Tl-2212 ref.\cite{Gerashenko1999}	                		&	102 (und)		&	16.08     	\\
	&		84	&	31.5		&&	112 (opt) 		&	17.35    	\\
	&		92	&	31.5		&&	104 (ovd)		&	17.87      	\\  \cline{4-6}
	&		93	&	31.5	& Tl-2223 Cu(1)			&	125 (opt) ref.\cite{Han1994}	 &	11.7		 \\ \cline{1-3}
Y-124 ref.\cite{haase_planar_2004} 		&	81	&	29.72  & Cu(2)	&		&	16.4		\\ \cline{1-3}
Y,Ca-123 ref.\cite{Williams2001}	&		68 (ovd) 	&	31.55	&	Cu(1)	&			115 (ovd?) ref.\cite{Zheng1996}	&	10.8		\\
		&		48 (ovd)  	&	31.65	&  Cu(2)	&				&	17.4		\\   \hline
Hg-1201	&			39 (und) ref.\cite{Gippius1997}	&	17.0	&Ba-0212F ref.\cite{Shimizu2011}	&			40 (und)	&	12.5\\ 
		&			45 (und) ref.\cite{RybickiU}	&	17.0	&&			73 (und)	&	13.7\\
		&			70 (und) ref.\cite{Gippius1997}	&	17.8	&&			105 (opt)	&	15.7	\\ 
		&			72 (und) ref.\cite{Gippius1997}	&	18.5  &&			102 (ovd)	&	16.5		\\ \cline{4-6}
		&			74 (und) ref.\cite{Rybicki2012b}	&	18.46  & CLBLCO\_0.1 ref.\cite{Keren2006} & 25 (und) & 30.6 \\
		&			95 (opt) ref.\cite{Gippius1997}	&	21.1	&& 41 (und) & 31.2\\
		&			96 (opt) ref.\cite{Gippius1997}	&	21.7	&& 58 (opt) & 31.8 \\
		&			97 (opt) ref.\cite{Rybicki2009}	&	20.88 && 21 (ovd) & 31.87	\\     
		&			89 (ovd) ref.\cite{Gippius1997}	&	23.1	&& 0 (ovd) & 31.93\\ \cline{4-6}
		&			85 (ovd) ref.\cite{RybickiU} 	&	22.4	  & CLBLCO\_0.4 ref.\cite{Keren2006} & 23 (und) & 27.1 \\
		&			0 (ovd) ref.\cite{Gippius1997}	&	27.2	  && 38 (und)  & 27.9 \\ \cline{1-3}
Hg-1212	&			125 (opt) ref.\cite{Ohsugi1996}	&	15.6	&& 75 (und)  & 28.3 \\ 
		&			117 (ovd) ref.\cite{Ohsugi1996}	&	16.4	&& 81 (opt) & 29.1 \\ \cline{4-6}
		&			101 (ovd) ref.\cite{Horvatic1994}	&	17.0	& Pr-214 ref.\cite{Jurkutat2014} & 0 & 12.2 \\
		&			101 (ovd) ref.\cite{Julien1996a}	&	17.6	&& 0 & 8.8\\ \cline{1-3}
Hg-1223 Cu(1)	&		115 (und) ref.\cite{Julien1996b}	&	9.7	&& 10 (und) & 4.8 \\ 
  Cu(2)	&				&	13.7	&& 25 (opt) & 0	\\ \hline
 Cu(1)	&			133 (opt) ref.\cite{Magishi1996a}	&	10.2 & Nd-214 ref.\cite{Jurkutat2014} & 0 & 14	\\ 
  Cu(2)	&				&	16.1		&& 25 (opt) & 1 \\ \hline
Hg,Cu-1223 Cu(1)			&	134 (opt) ref.\cite{Breitzke2004}	&	15.3	 &La -112  & 0 ref.\cite{mikhalev_63cu_1998} & 7.4 \\ 
  Cu(2)	&				&	16.6	&& 28 (und) ref.\cite{mikhalev_63cu_1998} & 3.6	\\ \cline{1-3}
Hg-1234 Cu(1)	&		85 (und) ref.\cite{Itohara2010}	&	9.3 && 40 (opt) ref.\cite{imai_superconductivity_1995} & $\le$4\\ 
  Cu(2)	&				&	14.8	&& 0 (ovd) ref.\cite{mikhalev_63cu_1998} & 2.3	\\ \hline
Cu(1)	&			123 (opt) ref.\cite{Itohara2010}	&	9.6	\\ 
  Cu(2)	&				&	17.8		\\  \cline{1-3}
Hg-1245 Cu(1)	&		108 (opt) Ref.\cite{Kotegawa2004}	&	8.37	\\ 
  Cu(2)	&				&	16		\\ \cline{1-3}
 \end{tabular}
\label{tab:two}
 \end{table}
\newpage

 \begin{table}
 \caption{Quadrupole splitting $^{17}\nu_\mathrm{Q}$ of planar oxygen for various families of cuprates in order of increasing doping ("und." - underdoped, "opt." - close to optimal doping, "ovd." - overdoped). For Tl- and Hg- families with three or more CuO$_2$ layers there are two planar O sites, the inner layer with O(1) and the outer layer O(2).
 \label{EFG-data}}
\small
  \begin{tabular}{| c | c | c | }
\hline
Family	&	$T_{\rm c}$ [K]		& $^{17}\nu_\mathrm{Q}$ [MHz]	\\ \hline \hline
La-214 ref.\cite{haase_planar_2004}	&	0	&		0.574		\\
  	&		22	&			0.6	\\
	&	38 (opt)	&	0.69		\\
	&	18	 &   0.81		\\   \hline
Y-123 ref.\cite{haase_planar_2004}	&	0		&		0.795	\\
	&		60	&	0.889	\\
	&		62	&	0.905	\\
	&		84	&	0.913	\\
	&		92	&	0.954	\\
	&	93 (opt)	&	0.986	\\
	&	93 (opt)	&	0.966	\\  \hline
Y-124 ref.\cite{haase_planar_2004} 	&		81	&		0.927	\\  \hline
Hg-1201  ref.\cite{Mounce2013}	&		74 (und)	&		1.050	\\ \hline																	
Tl-2201 ref.\cite{Kambe1993, zheng_local_1995}	&	85 (opt)	&		1.154	\\
	&	10 (ovd)	&	1.220	\\ \hline																	
Bi-2212 ref.\cite{Takigawa1994, Crocker2011}	&		86 (opt)	&	1.14	\\
                 		&	82 (ovd)	&	 	1.14	\\ \hline
TlSr-2212	&	102 (und) ref.\cite{Gerashenko1999}   		&	1.05			\\	
               		&	112 (opt) ref.\cite{Gerashenko1999} 		&	1.10			\\
              		&	104 (ovd) ref.\cite{Gerashenko1999}  	&	1.13\\
                     	&	103 (ovd) ref.\cite{Trokiner1995} 	&	1.09	\\  \hline
Tl-2223 O(1)	&		115 (ovd?)  ref.\cite{Zheng1996}&	1.06 \\ 
  O(2)	&			&	1.12	\\ \hline
Hg,Tl-1223  O(1)		&	132 (opt) ref.\cite{Lim1994}	&	1.101 \\ 
   O(2)		&	&					1.217 \\ \hline
CLBLCO\_0.1 ref.\cite{Amit2010} & 0 & 0.8 \\
& 0 & 0.89 \\
& 28 & 0.96 \\
& 36 & 0.98 \\
& 42 & 0.99 \\
& 55 (opt) & 1.02 \\ \hline
CLBLCO\_0.4 ref.\cite{Amit2010} & 0 & 0.88 \\
& 14 & 0.91 \\
& 36 & 0.94 \\
& 77 (opt) & 0.98 \\ \hline
Pr-214 ref.\cite{Jurkutat2014} & 0 & 0.905 \\
& 0 & 0.943\\
&10 & 0.933 \\
&25 (opt) & 0.921 \\ \hline
Nd-214 ref.\cite{Jurkutat2014} & 0 & 0.905 \\
& 23 (opt) & 0.885 \\ \hline

 \end{tabular}
\label{tab:one}
 \end{table}

\bibliography{EFG}{}
\bibliographystyle{apsrev4-1}